\documentclass[preprint,aps,physrev,amsmath,amssymb,superscriptaddress,nofootinbib,floatfix]{revtex4-2}

\usepackage{graphicx}
\usepackage{dcolumn}
\usepackage[usenames]{color}
\usepackage{setspace}
\usepackage{amsmath}
\usepackage{amsthm}
\usepackage{mathrsfs}
\usepackage{epsf}
\usepackage{epsfig}
\usepackage{bbm}
\usepackage{bm}
\usepackage{amsfonts}
\usepackage{amssymb}
\usepackage{latexsym}
\usepackage[english]{babel}
\usepackage{multirow}
\usepackage{float}
\usepackage{url}
\usepackage{slashed}
\usepackage{xcolor} 
\usepackage[utf8]{inputenc}
\usepackage{stmaryrd} 
\usepackage{enumitem}
\usepackage{siunitx}
\usepackage{verbatim}
\usepackage[colorlinks,linkcolor=black,anchorcolor=blue,citecolor=red,urlcolor=blue,]{hyperref}
\usepackage{etoolbox}

\begin{document}
\newcommand{\BH}{{\text{BH}}}
\newcommand{\BSM}{{\text{BSM}}}
\newcommand{\revt}[1]{{\color{black} #1}} 
\title{Self-interaction effects on the Kerr black hole superradiance and their observational implications}

\author{Ning Xie}
\email{xien27@mail2.sysu.edu.cn}

\author{Fa Peng Huang}
\email{Corresponding Author.  huangfp8@sysu.edu.cn}

\affiliation{MOE Key Laboratory of TianQin Mission, TianQin Research Center for Gravitational Physics \& School of Physics and Astronomy, Frontiers Science Center for TianQin, Gravitational Wave Research Center of CNSA, Sun Yat-sen University (Zhuhai Campus), Zhuhai 519082, China }

\begin{abstract}
Through the black hole (BH) superradiance, ultralight bosons can form dense clouds around rotating Kerr BHs. Certain ultralight bosons, such as axions and axion-like particles (promising dark matter candidates), naturally possess self-interactions, and thus may significantly modify the dynamics of the superradiance process. Previous studies on the detection or constraint of ultralight bosons through superradiance have usually neglected the self-interaction effects of bosons. In this work, we investigate the formation and evolution of self-interacting boson clouds in the full Kerr spacetime during BH superradiance. Using numerical methods, we compute the superradiant growth rate of boson clouds with self-interactions around Kerr BHs and quantitatively evaluate how the self-interaction strength of scalar bosons affects the growth rate. We also assess the evolution of the BH's mass and spin. Our results reveal that, in addition to the superradiance-imposed upper bound on the boson cloud mass, self-interaction of ultralight bosons introduces a new, lower critical mass limit, beyond which the growth rate of the boson cloud approaches zero. This implies that the superradiance process terminates earlier when self-interaction is considered. Furthermore, we explore how self-interaction affects both the oscillation frequency of boson clouds in gravitational atoms and the frequency of gravitational wave (GW) emitted through cloud annihilation. The anticipated frequency shift might be detectable by the GW observatories. Given that self-interaction substantially alters the evolution of BH superradiance, its effects can significantly relax existing constraints on scalar bosons derived from superradiance. Taking the spin measurements from GW190412 and GW190517 as examples, we discuss the impact of self-interaction on constraint results in details.
\end{abstract}

\maketitle

\section{Introduction}
The ultralight bosons, such as the axions~\cite{a1,a2,a3,a4,a5,a6,a7,Sikivie:2020zpn} and axion-like particles~\cite{Witten2006}, naturally correlate particle physics, cosmology, and gravity in many important processes, one of which is the superradiance mechanism~\cite{1971SR,1972SR,1973SR,Zouros:1979iw,Khlopov:1985fch,
Dolan:2007mj,2020SR} of Kerr black hole (BH). These fundamental particles can serve as the well-motivated dark matter  candidates~\cite{CDM,CDM2,CDM3,Adams:2022pbo}. The superradiance process can amplify bosonic waves near a fast-spinning BH by extracting energy and angular momentum of BH and the macroscopic condensed boson cloud can form spontaneously. The boson cloud consisting of ultralight boson particles exhibits an exponential growth due to the boundary condition at the outer horizon of Kerr BH. The superradiant instability of ultralight bosons can be significant when the Compton wavelength of the boson particle is comparable to the BH size, and consequently, the mass of boson cloud could reach up to $\sim10\textrm{\%} $ of the BH mass after the growth saturates. The BH–boson cloud system, so-called gravitational atom, can form for either the isolated BH or the BH in compact binary.  

BH superradiance~\cite{1971SR,1972SR,1973SR,Zouros:1979iw,Khlopov:1985fch,Dolan:2007mj,2020SR}, which minimally only requires the gravitational interaction of the ultralight bosons to be induced, is a well-established probe in the search for the ultralight bosons and dark matter. 
The formation of the dense, macroscopic bosonic condensate consisting of ultralight particles around the BH can lead to various observable signatures. These effects caused by the superradiance include the modification on the dynamic of BH inspiral~\cite{Arvanitaki:2016qwi,Zhang:2018kib,Zhang:2019eid,Hannuksela:2018izj,Baumann:2018vus,Baumann:2021fkf,Baumann:2022pkl,Xie:2022uvp}, the continuous GW emission from the boson cloud~\cite{Arvanitaki:2010sy,Arvanitaki:2014wva,Brito:2017wnc,Brito:2017zvb,East:2018glu,Siemonsen:2022yyf,Yang:2023vwm,Yang:2023aak}, and the spin-down of BH~\cite{Arvanitaki:2009fg,Witek:2012tr,East:2017mrj,East:2017ovw,Ng:2019jsx,Ng:2020ruv,Du:2022trq}. 

The discovery of gravitational waves (GWs)~\cite{LIGOScientific:2016aoc}  has initiated new perspectives to explore the dark matter  through various GW experiments including LIGO~\cite{aLIGO2010}, TianQin~\cite{TianQin:2015yph,TianQin:2020hid,TianQin:2025sos}, LISA~\cite{LISA:2017pwj}, and Taiji~\cite{Hu:2017mde}.
In order to investigate the signatures from the superradiant growth of ultralight bosons, it is important to rigorously understand the formation and dynamical evolution of the boson cloud, and a detailed evaluation of the BH–boson cloud system is required to make the appropriate interpretation for the current and upcoming observations and the constraints on the boson parameter space from the experiments.  

Unlike in the purely gravitational scenario for single BH–boson cloud system, the evolution of the gravitational atom can be influenced significantly due to the self-interactions of the scalar bosons~\cite{Baryakhtar:2020gao,Omiya:2022gwu}, other interactions with Standard Model particles~\cite{Rosa:2017ury,Sen:2018cjt,Ikeda:2018nhb,Mathur:2020aqv,Spieksma:2023vwl,Chen:2023vkq}, and the tidal interaction from the macroscopic companion object~\cite{Baumann:2018vus,Baumann:2021fkf,Baumann:2022pkl,Choudhary:2020pxy,Tong:2022bbl,Fan:2023jjj,Cannizzaro:2023jle,Boskovic:2024fga,Guo:2024iye}. Effects of self-interactions on the boson cloud evolution have attracted increasing attention~\cite{Baryakhtar:2020gao,Omiya:2022gwu, Chia:2022udn,Witte:2024drg,Lambiase:2025twn} and become important in the experimental searches for ultralight bosons~\cite{Palomba:2019vxe,Sun:2019mqb,Ng:2020ruv, LIGOScientific:2021rnv,LIGOS:2022osp}. 
Ref.~\cite{Baryakhtar:2020gao} addresses these questions with a detailed treatment of scalar self-interactions and considers the full dynamical evolution of a gravitational atom. Earlier than that, some of the studies argue during the growth of boson cloud, the strong attractive self-interactions would finally lead to the nonperturbative collapse of the cloud and consequently the drastic phenomenon called ``bosenova''~\cite{Arvanitaki:2009fg,Yoshino:2012kn,Yoshino:2013ofa}. Nonetheless, the current study~\cite{Baryakhtar:2020gao} proposes that when considering the energy exchange and level mixing between different bound states of the gravitational atom, the quartic self-interaction will cause the boson cloud to saturate at a new equilibrium with less occupation number. As a result, there is low probability for the bosenova to occur in the parameter space of interest. Overall, although analyzed at the semi-analytic level, this study~\cite{Baryakhtar:2020gao} demonstrates that the existence of  self-interaction for bosons can dramatically alter the dynamical evolution of BH superradiance. The boson cloud can stay at a two-level equilibrium stage and the evolution of BH parameters is modified. In Ref.~\cite{Baryakhtar:2020gao}, the spectra of boson clouds in the gravitational atom are described by the hydrogenic modes in the non-relativistic approximation, and the results are widely valid in the regime where the spacetime curvature at the position of boson cloud can be neglected.   

In contrast to Ref.~\cite{Baryakhtar:2020gao}, Ref.~\cite{Chia:2022udn} investigates the effects of the self-interactions on the saturated boson cloud in the nonlinear regime. In Ref.~\cite{Chia:2022udn}, the authors solve for the saturation stage of the gravitational atom for the dominant bound state considering the backreaction of the cloud on the metric, but ignore the effect of BH spin so as to simplify the spacetime to be spherically symmetric. 
The authors propose that, in comparison to the case without self-interaction, the mass of the self-interacting boson cloud in the saturated stage may increase by up to 70\% when the coupling parameter lies within $-10^{-2} \lesssim \lambda \lesssim 10^{-2}$.
Nonetheless, the additional backreaction effect of boson cloud on the metric is always negligible.
Also, the authors find an approximate quadratic scaling relation between the mass of the boson cloud and the coupling parameter. 
However, for the ultralight bosons, if we consider a much smaller coupling parameter for which the energy scale of the decay constant is around the Grand Unified Theory scale or the Planck scale, the increment of the cloud mass due to the effects discussed in~\cite{Chia:2022udn} would be insignificant. As in Ref.~\cite{Baryakhtar:2020gao}, we will focus on the case in which the energy scale of the decay constant for the ultralight bosons larger than $ 10^{13}~\mathrm{GeV}$ is considered in this work, since even a small value of self-interaction parameter can play an important role in the formation and evolution of boson cloud before it is \textit{saturated}. 

We investigate the evolution of the gravitational atom from the initial stage that the boson cloud formation begins to the cease of cloud growth, taking into account the self-interaction effects on the dominant bound state of boson cloud. We solve the field equation of scalar bosons in the vicinity of BH and find the eigenfrequency for the dominated bound state on the full Kerr metric. We then analyze in detailed the growth rate of self-interacting boson cloud and the evolution of BH parameters. Our calculation is reliable for a relatively large decay constant of boson field and under the assumption that the growth rate of boson cloud is much slower than the oscillation frequency of boson field around the BH. In fact, we find that over a large range of parameter space of interest for the gravitational atom, these approximations are sufficiently valid. We calculate the results within a relativistic framework. Thus, the results can be applied to the gravitational atom with a large range of its parameters such as BH mass and boson mass. It is necessary to note that, if we adopt the non-relativistic approximation, the treatment of the self-interaction term for a single-state boson cloud in this study  only leads to a quantitative shift in the energy potential of the bound state, as also analyzed in Ref.~\cite{Baryakhtar:2020gao}.  When we solve the boson field equation in the flat-spacetime limit, the deviation of the boson cloud growth rate disappears . Therefore, the modification on the growth rate of the boson cloud comes from not only the self-interactions of the bosons, but also the curvature nature of spacetime background.   

The paper is organized as follows. We first review the well-known superradiance process and introduce the self-interaction of scalar boson. We focus on the evaluating method to impose the quartic self-interaction term for scalar particles in Sec~\ref{met}. In Sec~\ref{rts}, we apply the numerical approach to compute the superradiant growth rate of self-interacting boson clouds around the BH. Then, we evaluate how the self-interaction among bosons alters the GWs emitted by the boson clouds in Sec~\ref{mod}. In Sec~\ref{csq}, we study in detail the potential impact of self-interactions on the existing observations and experiments, especially the relaxation of constraints on scalar bosons. We conclude and discuss future directions in Sec.~\ref{cc}.   
\section{\label{met}Superradiance and boson cloud with self-interaction}
In this section, we present a concise overview of the superradiance phenomenology for a (free) scalar field and especially the quartic self-interaction term of the scalar field.  We then focus on the numerical method to compute the spectra of a gravitational atom considering the self-interaction of boson clouds. 
We adopt the units in which $ \hbar= G =c=1 $, unless otherwise specified. 
\subsection{Black hole  superradiance}
The boundary condition at horizon of the Kerr BH leads to imaginary frequency of ultralight particles and then the exponential growth of the bound-state scalar bosons. In other words, a rapidly rotating BH can spontaneously transform energy and angular momentum to generate a macroscopic scalar cloud and form a gravitational atom through superradiance process~\cite{1971SR,*1972SR,1973SR,Zouros:1979iw,Khlopov:1985fch,Dolan:2007mj,2020SR}. 
In quantum field theory, the dynamics of a scalar boson field in the Kerr spacetime are governed by the Klein-Gordon equation. 
In general case, solving this equation numerically yields the eigenstates and the frequency eigenvalues of the scalar cloud. It has been demonstrated that the boson field exhibits a positive imaginary component in its frequency eigenvalue. This positive imaginary part results in an exponentially growing solution for the boson cloud when the Compton wavelength of the boson is comparable to the scale of the BH. Nevertheless, up to date, the numerical studies on Klein-Gordon solution of boson cloud's formation on the Kerr background has not taken into account the self-interaction of the field.

\revt{
For the scalar boson, the Lagrangian of the field $\phi$ on the Kerr metric is 
\begin{equation}\label{lagsf}
    \mathcal{L}=-\frac12(\nabla_\mu\phi)(\nabla^\mu\phi)-V(\phi)  \,\,,
\end{equation}
where $V(\phi)$ is the effective potential at low energy scale. 
For a free boson field, the effective potential takes the form 
\begin{equation}V_{\rm free}(\phi) =\frac{1}{2}{m_a ^2}\phi^2 \,\,.\end{equation}
Therefore one can have the Klein-Gordon equation 
\begin{equation}(\nabla^2-m_a^2)\phi=0\end{equation}
for the free field. The metric convention here is $(-,+,+,+)$. 
}

In the gravitational atom, the gravitational fine-structure constant is defined as $ \alpha =  { M_\BH} m_a $, where $ m_a $ is the mass of scalar boson and $M_\BH$ is the Kerr BH mass. 
\revt{This parameter plays a critical role in determining  the efficiency of the superradiant instability and the time scale of scalar cloud formation.  For the scalar field in the vicinity of the Kerr BH, the boundary conditions at horizon and infinity lead to imaginary eigenfrequency of the bosons.  
The boson cloud with real eigenfrequency $\omega_\text{Re}$ and magnetic quantum number $m$ could extract energy and angular momentum from BH spontaneously, if the initial spin of BH is greater than the critical value that triggers the superradiance process with the following superradiance condition~\cite{Dolan:2007mj,Arvanitaki:2014wva,2020SR} reads 
\begin{equation}\label{SR}
	\omega{}{}_\text{Re} < \dfrac{m\chi}{2r_{+} } \,\,,
\end{equation}
where $ \chi $ is the dimensionless spin of rotating BH ($0<\chi<1$) and $r_{+}={M_\BH}\left( 1+\sqrt{1-\chi^{2}} \right)  $ is the radius of the outer event horizon. 
Although in some previous references this condition was derived analytically, it could also be demonstrated by numerically solving the superradiant instability rate, see Fig.~\ref{figsbh} for details. 
Since a significant  cloud growth rate requires $ \alpha \sim \mathcal{O}(10^{-2}) \textrm{--}\mathcal{O}(  1) $, for the mass range of the astrophysical BHs, we are able to investigate the boson mass ranging from $ 10^{-21} $ to $ 10^{-10}~\text{eV} $. In this work, for simplicity, we choose the benchmark BH mass $M_\BH = 25,\ 40,\ 60,\text{and } 100~{\rm M_\odot}$, and focus on the boson mass range $ 10^{-14}\,\textrm{--}\, 10^{-12}~\text{eV} $.}

The superradiance rates-----imaginary parts of bound state frequencies-----depend on boson mass and BH spin. 
Specifically, for $ 2p $ level bound state ($\, \left|211\right\rangle$ state), the boson cloud grows and saturates with the timescale of  $ \tau_\text{SR} =1~\text{day} \left(\frac{ M_\BH}{40~{\rm M_{\odot}}}\right) \left(\frac{{ 0.2}}{_{\,}{ \alpha}_{\,}}\right)^9 \left(\frac{ 0.9}{{ \chi}}\right) $~\cite{Brito:2017zvb}. 
When exponential growth stops the scalar cloud extracts part of angular momentum from BH, and the \revt{occupation} number of scalar boson could be evaluated by~\cite{Arvanitaki:2014wva}
\begin{equation}\label{Naxioncloud}
	N_{a\rm{}} \sim 10^{77}
	\left(\dfrac{\varDelta \chi}{0.1}\right)\dfrac{1}{m}\left(\dfrac{M_\BH}{40~{\rm M_{\odot}}}\right)^{2}\,\,,
\end{equation}
with $ \varDelta \chi $ the spin difference \revt{of the BH from the initial stage when the boson cloud formation begins. Boson cloud formation depletes BH spin through continuous transfer of angular momentum from BH to the bound state. Eq.~\eqref{Naxioncloud} demonstrates the conservation of angular momentum.}

When ${\alpha}/{m}$ is much smaller than unity, the angular frequency eigenvalue of bound state can be solved perturbatively. If assuming the boson cloud to be non-relativistic, we can factor out the phase term in the field function $\phi$ related to its mass, and the remaining scalar function satisfies the Schrödinger-like equation. 
By solving the equation, one can in principle determine the eigenstate and (real) frequency eigenvalue of scalar bosons analytically.  
The eigenstates of boson clouds are analogous to hydrogenic bound states in the non-relativistic limit. 
On the other hand, 
in the limit that $\alpha$ is much smaller than unity, one can employ the Detweiler's approximation in order to derive the imaginary frequency of boson clouds, and give the analytical expressions of the growth rates~\cite{Detweiler:1980uk,Baumann:2019eav}.  

However, as we have mentioned, in principle the superradiant growth rate has to be calculated numerically for the general cases in which $\alpha $ could be relatively large. 
Besides, the nonlinear self-interaction term also brings extra difficulties in applying the Detweiler's approximation to determine the growth rate of boson clouds with the analytical approaches. 
The analytical approaches typically require the simplifications of the boson field equation 
which are only valid in the non-relativistic approximation. 
As for the relativistic case, we introduce the numerical method to solve the Klein-Gordon equation in the Kerr spacetime in the next subsection. 

In the following of this section, we numerically solve for the quasi-bound state spectrum of the scalar cloud around the Kerr BH, without taking the assumption that ${\alpha}$ should be much smaller than unity. 
Before introducing the self-interaction term of the scalar boson, we review the numerical method for finding the bound states of the  scalar field in Ref.~\cite{Dolan:2007mj}, and employ the continued fraction method to determine the scalar frequency eigenvalue. For a more detailed analysis about the technique, see also Ref.~\cite{Baumann:2019eav}. 
\subsection{Numerical solution for the ultralight scalar bosons}
The continued fraction method was used to find the  bound-state modes of the boson field. 
Since the Kerr background is axisymmetric and stationary, we assume the field function of boson cloud $\phi$ is separable and it is consisted of the radial part and the angular part and so on. Explicitly, the Kerr metric in Boyer-Lindquist coordinates is 
\begin{equation}\mathrm{d}s^2=-\frac{\Delta}{\Sigma}\left(\mathrm{d}t-{j}\sin^2\theta\mathrm{d}\phi\right)^2+\frac{{\Sigma}}{\Delta}\mathrm{d}r^2+{\Sigma}\mathrm{d}\theta^2+\frac{\sin^2\theta}{{\Sigma}}\left({j}\mathrm{d}t-(r^2+{j}^2)\mathrm{d}\phi\right)^2 \,\,,\end{equation}
where
\begin{equation}\Delta= r^2-2{M_\BH}r+{j}^2 \,\,,\end{equation}
\begin{equation}
\Sigma= r^2+{j}^2\cos^2\theta \,\,.
\end{equation}
Note that ${j} = \chi{M_\BH}$ is the BH angular momentum parameter\footnote{Here and in the following, we use $j$ to denote the BH angular momentum parameter  instead of $a$, thereby avoiding the confusion with other parameters.}. 
The solutions of $\Delta=  0$ represent the BH event horizons at 
\begin{equation}r_{\pm}={M_\BH}\pm\sqrt{{M_\BH}^{2}-{j}^{2}} \,\,.\end{equation}
For the scalar field, the radial and angular part of the field function $R(r)$ and $S(\theta)$ satisfy the confluent Heun equation and the spheroidal harmonic equation, 
\begin{equation}\begin{aligned}
\frac{d}{dr}\biggl(\Delta\frac{dR_{lm}}{dr}\biggr) & +\biggl[\frac{\omega^2(r^2+{j}^2)^2-4{M_\BH}{j}m\omega r+m^2{j}^2}\Delta \\
 & -(\omega^2{j}^2+{m_{a}}^2r^2+\Lambda_{lm})\biggr]R_{lm}(r)=0\,\,,
\label{KGr}\end{aligned}\end{equation}
\begin{equation}\label{KGa}
\frac{1}{\mathrm{sin}\theta}\frac{d}{d\theta}\left(\mathrm{sin}\theta\frac{dS_{lm}}{d\theta}\right)+\left[{j}^2(\omega^2-{m_{a}}^2)\mathrm{cos}^2\theta-\frac{m^2}{\mathrm{sin}^2\theta}+\Lambda_{lm}\right]S_{lm}(\theta)=0\,\,,\end{equation}
where $\omega_{}$ denotes the frequency eigenvalue of the scalar cloud, $n$ is the principal quantum number, $l$ is the total angular momentum, and $\Lambda_{}$ represents the angular eigenvalue of specific bound state\footnote{Hereinafter we eliminate the subscript which marks the quantum numbers for the brevity.}. Due to the spheroidicity, the eigenvalue $\Lambda_{}$ exhibits a non-trivial deviation, which can be expanded as a power series in ${j}^2(\omega^2 - {m_{a}}^2)$,
\begin{equation}\Lambda_{}=l(l+1)+\sum_{n=1}^\infty u_n\left[{j}^2(\omega^2-{m_{a}}^2)\right]^{2n} \,\,,\end{equation}
where the first several coefficients $u_n$ had been calculated by Seidel~\cite{Seidel:1988ue}. 
On the Kerr background, due to the behavior of the scalar field at the horizon and infinity, we can expand the radial function $R(r)$ as 
\begin{equation}R(r)=(r-r_+)^{-iP}(r-r_-)^{iP+Q-1}e^{qr}\sum_{n=0}^\infty a_n\left(\frac{r-r_+}{r-r_-}\right)^n\label{sffr}\end{equation}
where
    \begin{align}
        q&=\pm\sqrt{{m_{a}}^{2}-\omega^{2}} \,\,,\\
        P&=\frac{2r_{+}(\omega-{{j}m}/{2{M_\BH}r_+})}{r_{+}-r_{-}} \,\,,\\
        Q&=\frac{{m_{a}}^2-2\omega^2}{q}\,\,,
    \end{align}
and $a_n$ are the expansion coefficients. By substituting Eq.~\eqref{sffr} into Eq.~\eqref{KGr} there are the recurrence relations of $a_n$,
\begin{equation}\label{rr}\alpha_na_{n+1}+\beta_na_n+\gamma_na_{n-1}=0 \,\,,\quad n\in\mathbb{N} \,\,,\end{equation}
where 
\begin{equation}
    \begin{aligned}
        \alpha_{n}&=n^2+(c_0+1)n+c_0 \,\,,\\
        \beta_{n}&=-2n^2+(c_1+2)n+c_3 \,\,,\\
        \gamma_{n}&=n^2+(c_2-3)n+c_4 \,\,,
    \end{aligned}
\end{equation}
with 
\begin{equation}\begin{aligned}
 c_0=\ &1-2i\omega-\frac{2i}{b}\left(\omega-\frac{{j}m}{2}\right) \,\,, \\
  c_{1}=\ &-4+4i(\omega-iq(1+b))+\frac{4i}{b}\left(\omega-\frac{{j}m}{2}\right)  -\frac{2(\omega^2+q^2)}{q}, \\
  c_{2}=\ &3-2i\omega-\frac{2(q^2-\omega^2)}{q}-\frac{2i}{b}\left(\omega-\frac{{j}m}{2}\right), \\
  c_{3}  =\ &\frac{2i(\omega-iq)^3}{q}+2\left(\omega-iq\right)^2b+q^2{j}^2+2iq{j}m-\Lambda_{lm}-1 \\
 & -\frac{\left(\omega-iq\right)^2}{q}+2qb+\frac{2i}{b}\left(\frac{\left(\omega-iq\right)^2}{q}+1\right)\left(\omega-\frac{{j}m}{2}\right), \\
 c_{4}  =\ &\frac{(\omega-iq)^4}{q^2}+\frac{2i\omega(\omega-iq)^2}{q}  -\frac{2i}{b}\frac{(\omega-iq)^2}{q}\left(\omega-\frac{{j}m}{2}\right) \,\,, \\
b=\ &\sqrt{1-{j}^2} \,\,,
\end{aligned}\end{equation}
and the initial condition is $a_{-1}=0$.

These recurrence relations have the minimal solution which satisfies the continued fraction equation \cite{1892Pincherle,1967Gaut}
\begin{equation}\frac{\beta_0}{\alpha_0}-\cfrac{\gamma_1}{\beta_1-\cfrac{\alpha_1\gamma_2}{\beta_2-\cfrac{\alpha_2\gamma_3}{\beta_3-\cdots}}}=0 \,\,.\label{cfe}\end{equation}
Therefore, we can numerically solve the transcendental equation~\eqref{cfe} to obtain the frequency eigenvalue of quasi-bound state.  The truncation we choose in the continued fraction decides the error in the result. 
In particular, when gravitational fine-structure constant $\alpha$ becomes too small ($ \alpha \sim \mathcal{O}(10^{-2} ) $), we need to keep more term in the continued fraction for the convergence of the result.  
\subsection{Self-interacting scalar bosons}
Axions, or axion-like particles, are promising dark matter candidates.
The searches for ultralight scalars like axions to date have not adequately accounted for self-interactions among particles in the superradiance process. Self-interactions, that could significantly alter superradiance dynamics, are generically present for many scalar models, including axion-like dark matter candidates and string axions. Besides the QCD axion arising as the result of the Peccei-Quinn  symmetry broken at a high energy scale $f_{a}$, more general axion-like particles can be present in particular theory and are typically predicted to arise in string theory, 
in which the symmetry-breaking scale $f_{a}$ can fall in a wide mass range~\cite{Witten2006,Arvanitaki:2009fg}. 
From a particle physics perspective, the simplest interaction of a scalar field is the quartic self-interaction. 

As for the superradiant instability, in general, one would expect that relatively strong self-interactions of the bosonic particles can lead to steady-states with orders of magnitude smaller occupation numbers than the expectation from purely gravitational growth. It is also important to rigorously include the effects of self-interactions in searches for GW signals. 

\revt{Let us still begin with the scalar field Lagrangian (cf. Eq.~\eqref{lagsf}) on the Kerr metric. }
Self-interacting ultralight bosons generally arising through non-perturbative theory at some high energy scale give an effective potential. Any type of potential of the scalar field $\phi$ can be expanded to the lowest order with the mass term and the quartic self-interaction term. The strength of the mass term in the potential is proportional to the square of scalar mass, and the quartic self-interaction term proportional to the dimensionless self-interaction parameter $\lambda$. That is,
\begin{equation}V(\phi) =\frac{1}{2}{m_a ^2}\phi^2-\frac{\lambda}{4!}\phi^4+\mathcal{O}(\phi^6) \,\,.\end{equation}
Since the quartic self-interaction  is the most generic and important interaction term for small amplitudes, we focus on this self-interaction term and omit the higher order in this work. By introducing the quartic self-interaction term into the Lagrangian, the equation of motion becomes 
\begin{equation}
    (\nabla^{2}-m_{a}^{2})\phi+\lambda\frac{\phi^{3}}{3!}=0 \,\,.
\end{equation}
Note that we choose $\lambda>0$ to represent the attractive self-interaction. Without losing generality, the coupling parameter $\lambda$ can be defined as
\begin{equation}
    \lambda=\frac{m_{a}^{2}}{f_{a}^{2}} \,\,,
\end{equation}
so in the following we equivalently use the energy scale $f_{a}$ to quantify the strength of scalar self-interaction for convenience. Thus, we have 
\begin{equation}\lambda\approx10^{-82}\left(\frac{m_{a}}{10^{-13}~\mathrm{eV}}\right)^2\left(\frac{10^{19}~\mathrm{GeV}}{f_{a}}\right)^2.\end{equation}
Then rearranging the equation of motion, 
\begin{equation}\label{neom}
    \left[\nabla^{2}-m_{a}^{2}\left(1-\frac{\phi^{2}}{3! f_{a}^{2}}\right)\right]\phi=0 \,\,.
\end{equation}
To the lowest order of $\phi$, heuristically Eq.~\eqref{neom} can be considered as the new Klein-Gordon equation on the Kerr metric and the deviation comes from the modification of the scalar mass. By considering the $\phi^{2}$ part as the background field, we can investigate this mass modification. The main contribution stems from the dominating energy level of bound state which grows at the maximum speed and occupies much more than others. The particle occupation number of the fastest growing level $N$ is proportional to the square of its field amplitude ($N\propto|\phi|^{2}$). We separate this part individually and define $\Pi$ to quantify the contribution of bound state wave function except $N$, i.e. $N\Pi={\phi^{2}}/\left.{ f_{a}^{2}}\right.$. 
The quantity $\Pi$ is only determined by the distribution of the wave function and the scale of $f_{a}$. We are going to evaluate this modification in a more realistic situation for the superradiant growth. The realistic approach is based on the fact that for the gravitational atom's parameter space of interest the growth timescale of the scalar cloud is much longer than the oscillation timescale of the particles, ${\omega}^{-1}$ ($\approx m_{a}^{-1}$). Also, the cloud growth timescale is much slower than the timescale of the particles' orbital motion in the vicinity of the BH. For these reasons, we can perform the time- and spatial averaging when investigating $\Pi$ within one period and calculating the scalar mass deviation. Using similar notation as in Eq.~\eqref{KGr} and Eq.~\eqref{KGa}, we can then calculate $\Pi$ with 
\begin{equation}
\Pi=e^{t\Gamma_{\rm SR} } \frac{| \overline{R}|^{2}|\overline{S}|^{2}}{ f_{a}^{2}}            \,\,,
\end{equation}
where $\Gamma_{\rm SR}$ is the growth rate of bosons, $t$ is the time and the bars on the symbols denote the average over spatial variables. In the non-relativistic limit, we already know the analytical expressions of radial and angular functions. The radial dependence of function, $R$, becomes radial hydrogen wave function, while the angular dependence of function, $S$, becomes the spherical harmonics. Based on the more thorough calculation, although the analytical results provide a good approximation, we carry out the estimation within the full Kerr background. We can investigate the mean values of the functions around the cloud region. Note that the radial and angular functions both obey the normalization condition.  
As for the occupation number at a specific time, we can estimate $N$ through the spin difference of the  BH due to the conservation of angular momentum. When only considering the fastest growing level of bound state, the scalar cloud is dominated by the particles of single state and the extracted angular momentum from Kerr BH is stored in the same state. The occupation number of the fastest growing state is 
\begin{equation}\label{N}
	N_{\rm{}} =
	\left.{\varDelta \chi}\right.\dfrac{ M_\BH^{2}}{m}=\left.{\varDelta \chi}\right.\ \alpha\dfrac{ M_\BH^{}}{m m_{a}}\,\,,
\end{equation}
where $ \varDelta \chi \lesssim \mathcal{O}(0. 1) $. 
Owing to the hierarchy of growth rates for different superradiant states, the newly-formed BH is always dominated by one bound state of scalar boson, and higher states would take much longer time to grow to the saturation. Thus, we consider almost all bound-state bosons occupy the same energy level here. 

Therefore, we can define the \emph{modified} boson mass through  
\begin{equation}\label{mmass}
\widetilde{m}_{a}^{2}=\left(1-\frac{N}{3!}\Pi\right)  {m}_{a}^{2}  \,\,.
\end{equation}
It should be noted that the availability of the Eq.~\eqref{mmass} requires $N\Pi \ll 1$. As we will see in Sec.~\ref{rts}, this condition can always be held for the several important ground states with small quantum number even when the state is saturated and $N$ attains the maximum. The deviation can be considered perturbatively.  
The frequency eigenvalue can be determined by the boson mass, e.g. 
\begin{equation}\omega =\left(1-\frac{\alpha^{2}}{2n^{2}}+\mathcal{O}(\alpha^4) \right)\widetilde m_{a} \,\,.\end{equation}
\subsection{The spectra of the gravitational atom }
We can apply the modification on the recurrence relation Eq.~\eqref{rr} and obtain the new continued fraction equation for self-interacting bound state boson. While the deviation of boson mass induced by the self-interaction potential is expected to be much small, we could only keep the leading-order modification in the calculation.  To be explicit, we neglect all the $\mathcal{O}(\delta^2)$ terms, where the relative deviation $\delta$ is defined by 
$   \delta \equiv (\widetilde{m}_{a}^{2}-{m}_{a}^{2})/{m}_{a}^{2}$. Given the quantum state of the boson cloud and the parameters of the gravitational atom, we can solve the frequency eigenvalue of the cloud and find out the rate of superradiant instability for the self-interacting boson. 
\revt{In general, modifications to the energy spectra of a gravitational atom can be classified into two distinct types. First, changes of the \textit{imaginary} part of the complex eigenfrequency directly affect the rate of superradiant instability in the self-interacting boson cloud, governing the exponential growth of the bound-state occupation number. As we will show later, the self-interaction could potentially suppress the growth rate of boson cloud; and this effect will be discussed in detail in Sec.~\ref{rts}. Second, the modification to the \textit{real} part of the eigenfrequency induces energy-level shift of the boson cloud, which will be specifically analyzed in Sec.~\ref{mod}. 
}

A necessary condition for obtaining a reliable solution for the superradiance rate is that 
\begin{equation}\label{cond1}
|\delta|=\frac{1}{3!} N\Pi  \ll 1  \,\,.
\end{equation}
This condition ensures that the modification to the boson mass due to self-interaction is negligible compared to the particle's original mass. Therefore, although the continued fraction equation generally provides a solution, the condition \eqref{cond1} must hold for the bound-state cloud to ensure the solution is physical. At the time the fast-spinning BH is born, for example when two primary BHs merge to a new BH or when a supernova explodes, the occupation number of bound state bosons is negligible. At the beginning of boson cloud growth, the small particle number keeps that the condition \eqref{cond1} is valid all through, while the self-interaction effect of ultralight boson is also very insignificant. When the bound-state bosons are produced by extracting the angular momentum of BH and eventually form a macroscopic cloud, the self-interaction term and the modification on the boson mass would become more and more important in the superradiance process. As we can find out, the relative deviation $\delta \propto N$. 

\revt{    
It is worth to have more discussion on the conditions required to ensure the validity of the numerical methods introduced in this section. Firstly, we adopt the continued fraction method to solve both real and imaginary components of the boson eigenfrequency. The continued fraction equation is derived based on the Klein-Gordon equation on Kerr metric, so it is applicable for  different value of $\alpha$. Both the superradiant growth rate and the energy level of the boson cloud can be calculated numerically for the general cases in which $\alpha $ might  be relatively large.   Secondly, the treatment of the self-interaction term as the modification to the boson mass is valid only when the occupation number satisfies $\frac{1}{3!} N\Pi  \ll 1  $. As the occupation number grows exponentially during the boson cloud formation, perhaps the condition may no longer satisfy once the occupation number is sufficiently large. However, the self-interaction effects generally bring a new upper limit that restrict the growth of occupation number. When the occupation number approaches  to the new bound, the growth rate approaches to zero. With more detailed calculations, we find that for the parameter space of interest, the boson cloud growth always ceases much earlier  before the condition~\eqref{cond1} breaks down. That  indicates it is safe  to treat the self-interaction as the perturbative modification on boson mass when considering the newly-formed self-interacting boson cloud. 
Moreover, as for the drastic bosenova  effect~\cite{Arvanitaki:2009fg,Yoshino:2012kn,Yoshino:2013ofa}, current studies~\cite{Yoshino:2012kn,Yoshino:2013ofa} had performed the numerical simulation for the axion cloud growth and the results had indicated  during the cloud growth, the strong self-interactions would lead to the nonperturbative collapse of the cloud eventually. 
However, the upper bound of cloud occupation number also potentially prevents the bosenova collapse, cause the bosenova occurs only when the cloud occupation number exceeds the critical amount. If the  new upper limit of occupation number is small than the critical value needed to trigger bosenova, there is low probability for the bosenova to occur during the boson cloud formation.  
Nonetheless, the numerical approaches introduced in this section assume the whole boson cloud is dominated by the fastest growing state, and neglect the  contribution from other higher states. 
Ref.~\cite{Baryakhtar:2020gao} had proposed the moderate and large self-interaction can transfer the bosons to other higher states and several bound states are excited in the boson cloud. This work shows that  when considering the energy exchange and level mixing between different bound states of the gravitational atom, the quartic self-interaction will cause the boson cloud to saturate at a new multi-state equilibrium, while the original fastest-growing state will have a lower occupation number.  
Although the growth timescales of higher superradiant states are much longer than the fastest one, the self-interaction can excite the transition between these states. After other higher states are populated, they bring extra effects on the growth of the dominant state. 
Yet these effects are far beyond the scope of our work. 
In the following calculations, we omit the contribution from the multiple states except the self-interaction of the fastest growing state itself. 
}

\section{\label{rts}The  superradiance rate of  the self-interacting boson cloud}
In this section, we numerically compute the growth rate of self-interacting boson cloud. 
We utilize the methods introduced in the previous section to model the self-interaction effects of the boson cloud. In applying the continued fraction method, it is important to retain a sufficient number of modes to ensure sufficient numerical precision\footnote{Previous studies~\cite{Dolan:2007mj,Baumann:2019eav} have indicated that for a moderately small $\alpha$, the required mode number is on the order of a few hundred. In contrast, for very small $\alpha$, the required mode number increases substantially, exceeding one thousand. 
}.  For the BH–boson cloud system parameters in our calculations, we choose a fixed BH mass, such that  the gravitational fine-structure constant $\alpha$ is directly determined by the boson mass. We compute the superradiance rate for the self-interacting boson and compare it with that for the free boson. 
\revt{Due to the significant disparity in growth rates across different superradiant states, the newly-formed BH is always dominated by single bound state of scalar boson. Higher-order states exhibit substantially longer saturation timescales. 
Thus, we model the system with the approximation that virtually all bound-state bosons populate the single energy level which has the fastest growth rate, and neglect potential level mixing~\cite{Baryakhtar:2020gao}. 
For the parameter space of interest, we consider only $\, \left|211\right\rangle$ state in this section.}

In the following, we perform detailed analyses for different spins of BH (Sec.~\ref{sbh}), different scalar self-interaction scales (Sec.~\ref{asi}) and different occupation numbers of boson cloud (Sec.~\ref{onac}).    
 
\subsection{\label{sbh}Different spins of BH} 
\begin{figure}[!th]
    \centering
    \includegraphics[width=0.9\linewidth]{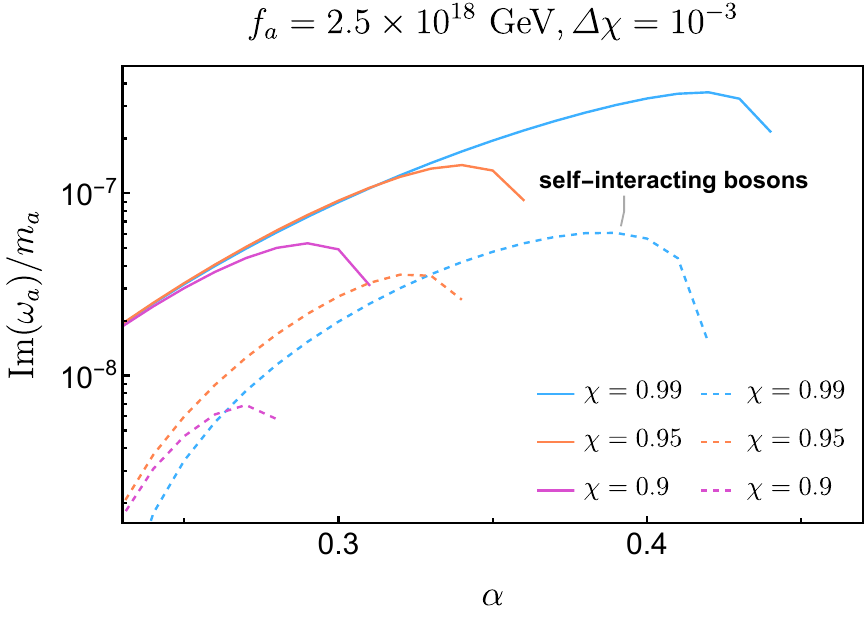}
    \caption{Growth rates of the self-interacting boson cloud for various BH dimensionless spins. The growth rates for the free boson are depicted with solid lines, while those considering the self-interaction term are shown with dashed lines. The benchmark values of  BH spin difference remain the same across  different lines. The growth rates are suppressed by approximately half an order of magnitude due to self-interaction. In all scenarios, the growth rate reaches a peak and then decreases as the $\alpha$ increases. }
    \label{figsbh}
\end{figure}

The spin of BH is the initial factor that triggers the superradiance process of the boson field. As the BH's angular momentum is transferred to the bound state during the formation of the boson cloud, the BH spin gradually decreases until the superradiance condition is no longer satisfied. When a new BH forms through binary BH mergers or supernova explosions, it is generally expected to possess a high initial angular momentum. Additionally, the BH spin can increase via astrophysical processes, such as disk accretion. The dimensionless spin of astrophysical BHs can reach values exceeding 0.998 during their evolution~\cite{Thorne:1974ve,Volonteri:2004cf,Sesana:2014bea,Izquierdo-Villalba:2020hfk,Reynolds:2019uxi,Burke:2020vvk}. We calculate the growth rate of scalar boson-----defined as the imaginary part of the frequency eigenvalue-----for a range of BH dimensionless spins $\chi=  0.9, 0.95, 0.99$. For the comparison between the self-interacting bosons and the free bosons, we plot the free boson results using solid lines, while the growth rates accounting for boson self-interaction are shown with dashed lines, as illustrated in Fig.~\ref{figsbh}.  

In Fig.~\ref{figsbh}, we choose a moderate value of self-interaction strength. The energy scale of self-interaction $f_{a}$ is around $2.5\times10^{18}~\mathrm{GeV}$. We use the same $f_{a}$ in the calculation for self-interacting boson clouds. As for the gravitational atom, although the BH spins we consider range from 0.9 to 0.99, the spin difference of the BH from the initial stage when the boson cloud formation begins remains unchanged. This approach ensures that the boson cloud maintains an almost equal occupation number. We specify the benchmark parameters on the top of each figure.

From the results of Fig.~\ref{figsbh}, we find that for the self-interacting boson cloud the growth rates become smaller owing to the  existence of self-interaction effect. The growth rates decrease by a half to one order of magnitude compared to the calculation results without self-interaction. In the free scalar field case, every solid line reaches a peak and then drops eventually as the $\alpha$ increases. For the self-interacting boson, the results have similar behavior. 
\revt{We can see that around these highest points, with the increase of $\alpha$, the growth rates decline and approach to zero. These critical values  of  $\alpha$ correspond to the superradiance condition for different BH spin. The allowed range of $\alpha$ for the superradiant growth of boson cloud is dependent on the value of $\chi$. For the larger BH spin, the boson cloud can grow via the superradiance process for the larger $\alpha$. These results can be understood from the superradiance condition~\eqref{SR}.
Moreover, when considering the self-interaction effects the upper bounds of the $\alpha$ ranges for each BH spin  remain approximately  unchanged.} That means in the case with self-interaction  the superradiance condition  would be still applicable.    

\subsection{\label{asi}Different  self-interaction scales} 
\begin{figure}[!t]
    \centering
    \includegraphics[width=0.8\linewidth]{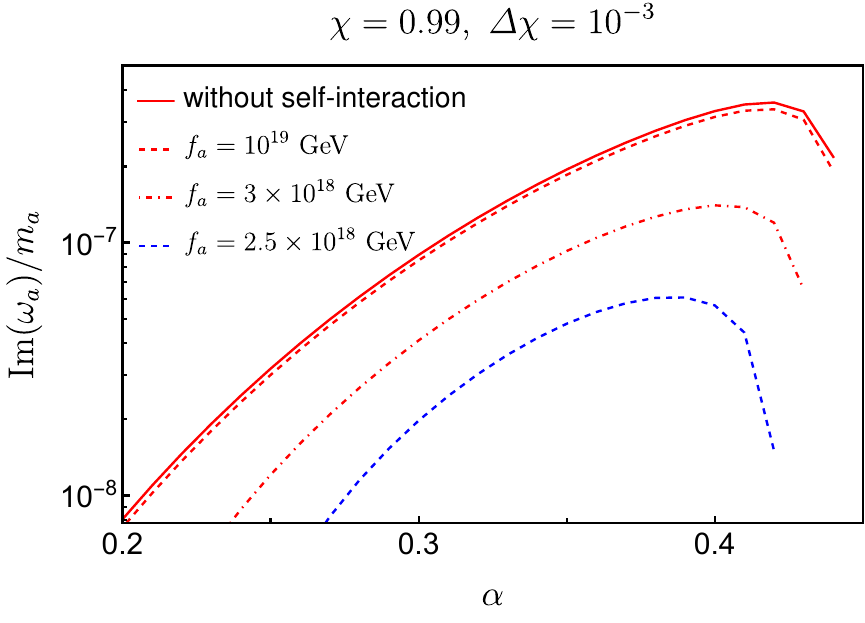}
  
    \caption{Growth rates of the self-interacting boson cloud for different values of $f_a$ and $\alpha$. The growth rate for a free boson is shown by the red solid line, while results for various $f_a$ values are represented by the red dashed line, red dot-dashed line, and blue dashed line. The spin and spin difference of the  BH are the same for all lines. The growth rate is suppressed by approximately half an order of magnitude when $f_a$ is less than $2.5 \times 10^{18}$ GeV.
    }
    \label{figasi1}
\end{figure}

We use self-interaction energy scale $f_{a}$ to describe the strength of quartic self-interaction. As the scale $f_{a}$ decreases, the self-interaction effect will be more significant. We compute the growth rate of self-interacting boson cloud for different  sets of  $f_{a}$ and $\alpha$. We consider the stage of the cloud growth that   the spin and the spin difference of  BH  are both the same. The result is shown in Fig.~\ref{figasi1}.

We observe that the deviations of the growth rate from the free boson case become more pronounced as the scale $f_a$ decreases, for the different values of $\alpha$ considered here. The growth rate is suppressed by approximately half an order of magnitude when $f_a$ is less than $2.5 \times 10^{18}$ GeV. However, for larger $f_a$, the growth rate closely approximates the result for a free boson (solid line). In fact, with more detailed calculations, we find that when $f_a$ is comparable to or greater than the Planck scale, the modification to the growth rate is generally negligible. For moderate self-interaction strengths, it would take a longer time for the boson cloud to grow, since the superradiance rate is smaller.     

\subsection{\label{onac}Different  occupation numbers of boson cloud} 
\begin{figure}[!t]
    \centering
    \includegraphics[width=0.8\linewidth]{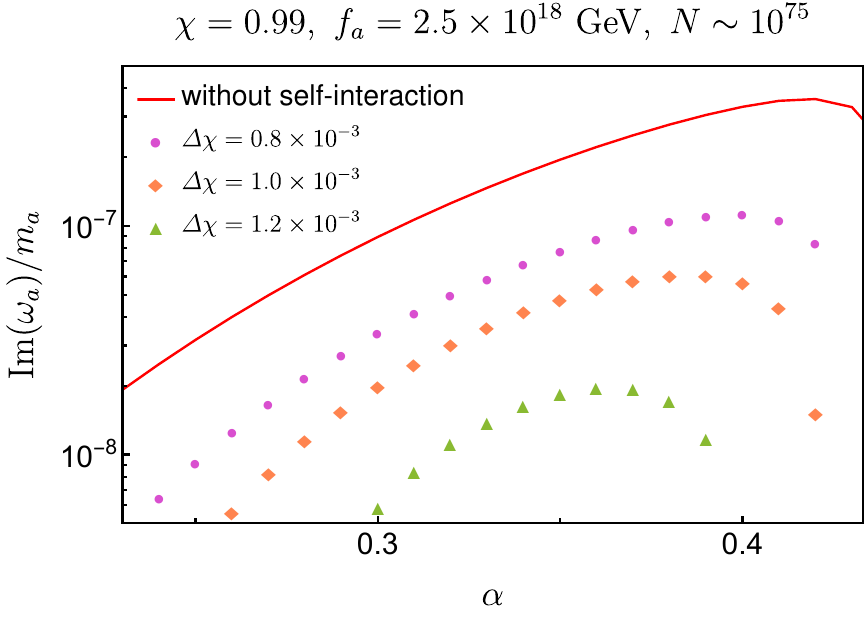}
    \caption{Growth rates of self-interacting boson cloud for different stages in the boson cloud amplification. The growth rates for the free boson are shown in red solid line, while the results when considering the self-interaction term are shown with data points.
    The BH spin and  self-interaction strength are the same for different cases. We find that the suppression on the growth rate becomes significant following the increase of the spin difference. }
    \label{figonac}
\end{figure}

In different stages of boson cloud formation, the occupation number increases exponentially until the superradiance condition is broken and the growth rate approaches  zero. By considering the particular BH spin and  self-interaction strength, we select a group of occupation numbers for the boson cloud. Since the BH spin difference $ \varDelta \chi $ in the superradiance process is directly related to the occupation number of fastest growing level in the boson cloud   via the conservation of angular momentum, we can equivalently use the occupation number $N$ as well as the $ \varDelta \chi $ to represent the different cloud formation stage.  We investigate the self-interaction effect for different $ \varDelta \chi $ and compute the growth rate of boson particles in the superradiance. We plot the result as the function of $\alpha$ in Fig.~\ref{figonac}. 

We find that the existence of cloud and its self-interaction suppresses the growth of cloud itself. This suppression on the growth rate becomes significant following the increase of the occupation number (spin difference). For a fixed $\alpha$, we can foresee that, in the superradiance process, the occupation number increases and the growth rate slows down. In fact, the growth rate can become zero and even negative finally as the occupation number increases. We should notice that this should be distinguished from the case of superradiance saturation and we can define the critical occupation number at which the growth rate approaches  zero on account of the self-interaction term. We denote this critical occupation number as $\widetilde{N}_{c}$ and the saturation occupation number in the superradiance condition as ${N}_{c}$. In general, we have 
\begin{equation}
    \widetilde{N}_{c} \ll {N}_{c}\,\,.
\end{equation}

\revt{
Before moving to the next section, we calculate the relative errors when solving for the superradiance rates of the self-interacting boson clouds. The relative error partly comes from the truncation we choose when solving the transcendental equation~\eqref{cfe}. We analyze the number of the terms in the continued fraction required for the convergence of solution.  We define the relative error criterion to determine the convergence of eigenfrequency as 
\begin{figure}[!h]
    \centering
    \includegraphics[width=0.8\linewidth]{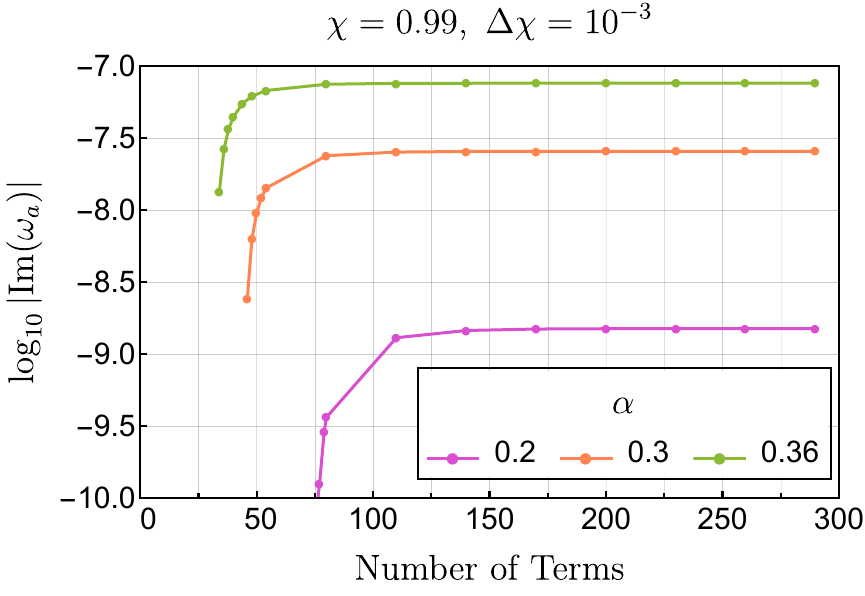}
    \caption{\revt{
    The imaginary part of eigenfrequency obtained from the numerical method, for different numbers of terms kept in continued fraction.  For the smaller $\alpha$, the solution converges more slowly and more terms in the continued fraction are required. The results are for the benchmark parameters, but similar conclusion holds for a wide range of parameter space. }
    }
    \label{figer}
\end{figure}
\begin{equation}
    \epsilon_n\equiv\left|\frac{\omega^{(n_{})}-\omega^{(n_{}-50)}}{\omega^{(n_{})}}\right| <10^{-4},
\end{equation}
where $\omega^{(k)}$ denotes the imaginary part of frequency eigenvalue computed with $k$ fraction terms. We show the result of imaginary eigenfrequency with different mode number in Fig.~\ref{figer}. It can be seen from Fig.~\ref{figer} that it takes more terms to be kept to ensure the convergence of the solution for a smaller value of $\alpha$. 
In the calculations of  this section, we choose the term number $\sim1000$ to ensure the convergence while optimizing computational efficiency.

 }

\section{\label{mod}The modifications of   self-interaction  on gravitational waves from gravitational atom}
In this section, we investigate how the quartic self-interaction of boson cloud alters the GWs radiated by gravitational atom. In addition to the cloud's quartic self-interaction, we  include the  self-gravity  of boson cloud as well. The analysis is mostly based on the results obtained in the previous section. The GWs are expected to be modified through two aspects. On  one hand, due to the self-interaction effect, the reduced saturation mass of boson cloud restricts the GW radiation power of boson cloud. On the other hand, the quartic  self-interaction and the self-gravity change the potential of the bound state when their strengths are comparable with the gravity of central BH, and the angular frequency of boson in the cloud, which determines the frequency of GWs sourced by the cloud annihilation, is different. 
\subsection{GW strain} 
In this subsection, we focus on the strength of the GW signals from the pair annihilation process in the boson cloud. We are interested in the GW radiation process after the cloud saturation because the boson occupation number has reached the maximum and the GWs have also reached its peak emission power. The GW radiation process could last for a much longer  timescale than the cloud formation. 
\revt{
During that period, the cloud depletes gradually if only considering the annihilation of bosons to GWs and the occupation number of $\, \left|211\right\rangle$ state evolves as 
\begin{equation}N(t)=\frac{N(t=0)}{1+t/\tau_\text{DC}}\,\,,\label{ONevo}\end{equation}
where $\tau_\text{DC}=(\Gamma_{a}N(t=0))^{-1} $ is the cloud depletion timescale through annihilation and is longer than $\tau_\text{SR} $ by a few of orders. $\Gamma_{a}$ is the annihilation rate of bosons, and $t=0$ corresponds to the onset of boson cloud saturation.   
Generally, the altered maximum occupation number of bound state boson is smaller than ${N}_{c}$ in the parameter space of interest  when introducing the self-interaction, as shown in Sec.~\ref{onac}. 
If considering a stronger coupling (smaller $f_{a}$), the maximum number of $N$ is suppressed more significantly.  
In the depletion process of cloud, the GW radiation power is proportional to $  N^2(t)$, so the GW amplitude received by the detector $h(t)$, is proportional to $N(t)$. On account of the scalar self-interaction, the maximum occupation number for the newly-formed boson cloud is $\widetilde{N}_{c}$. Thus, the peak GW amplitude is suppressed by the factor $\widetilde{N}_{c}/{N}_{c}$ (cf. Sec.~\ref{onac}). 
In Sec.~\ref{rts}, we has proposed that $\widetilde{N}_{c}$ is usually much smaller than ${N}_{c}$. In that respect, whether we consider the self-interaction effect could alter the peak GW amplitude significantly due to the change of cloud occupation number. 
\begin{figure}[!t]
    \centering
    \includegraphics[width=0.88\linewidth]{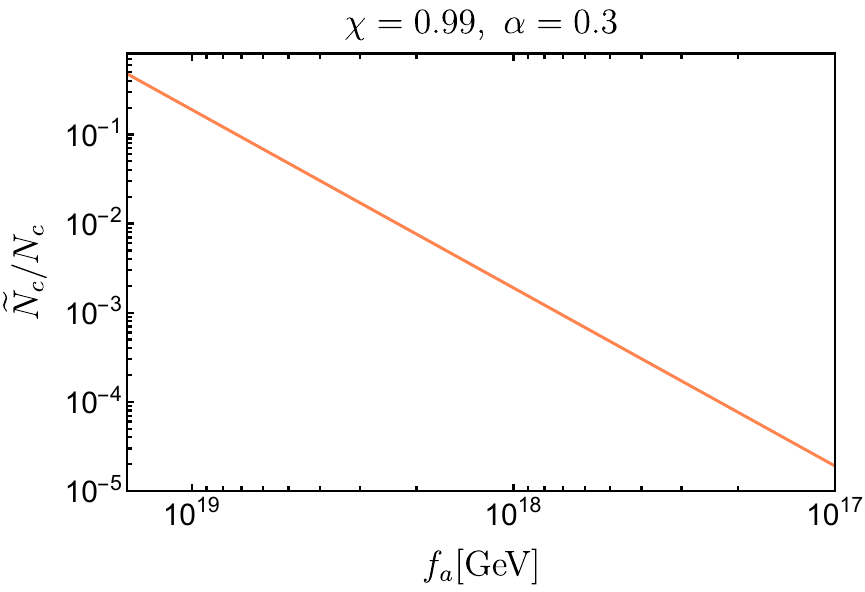}
  \revt{\caption{The suppression factor of peak GW amplitude for the continuous GWs emitted by the self-interacting boson cloud. }
    \label{figsurfa}}
\end{figure}
In Fig.~\ref{figsurfa}, we demonstrate the GW amplitude suppression factor $\widetilde{N}_{c}/{N}_{c}$ as the function of the self-interaction scale $f_{a}$. It can be seen from the plot that, for the smaller self-interaction scale, the maximum occupation number of boson cloud would decrease significantly, leading to a lower peak GW strain $h$.
Moreover, because of the GW radiation, the boson cloud will progressively lose  its energy after its saturation and its occupation number will decrease with time\footnote{This is always correct no matter whether the self-interaction is included. The  difference when introducing the self-interaction effect is the alteration of occupation number decreasing rate.}. 
In contrast to the  newly-formed boson cloud, for the boson cloud at a given age, the emitted GW strain will also become smaller with time. 
It can be seen from Eq.~\eqref{ONevo}  that, if the boson cloud has evolved with the age much longer than $\tau_\text{DC}$, its occupation number could be estimated as $N(t)\sim{\tau_\text{DC}N(t=0)}/{t} =(\Gamma_{a}t)^{-1}$. 
Therefore, when the age of boson cloud is much longer than $\tau_\text{DC}$, the initial maximum occupation number $N(t=0)$ (which is equal to $\widetilde{N}_{c}$) becomes irrelevant and  the GW amplitude $h$ roughly only depends on the annihilation rate to graviton $\Gamma_{a} $ and the time $t$. In this case, the concrete value of $f_{a}$ seems not so crucial. However, since the GW amplitude $h$ has become very small, this circumstance is out of interest.    
}
\subsection{GW frequency shift} 
\begin{figure}[h]
    \centering
    \includegraphics[width=0.8\linewidth]{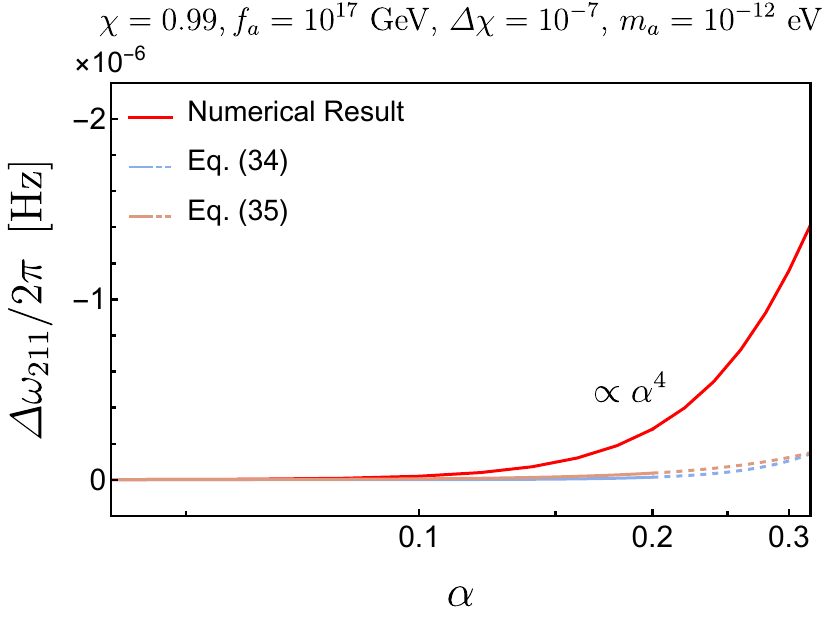}
    \caption{The frequency shift of the bound state calculated with numerical method within the Kerr framework. When $ \alpha $ is small, the numerical  result agrees well with the flat-spacetime estimation, Eq.~\eqref{freqq}.  }
    \label{figgw1}
\end{figure}
Using the method in Sec.~\ref{met}, we can compute the frequency eigenvalue of self-interacting boson. We compare the real part of boson frequency with and without considering the self-interaction coupling. Consequently, we can obtain the boson frequency shift which comes from the particles' self-interaction. This method can calculate the frequency shift in the particular stage of self-interacting cloud growth numerically, without any assumption on the background. Nevertheless, in the non-relativistic limit, the frequency shift can be evaluated analytically,  if given the boson cloud mass. Following Ref.~\cite{Baryakhtar:2020gao}, the frequency shift for the $\, \left|211\right\rangle$ state is approximately 
\begin{equation}\label{freqq}\varDelta \omega_{211}= -1.2\times10^{-4}\frac{\alpha^5}{M_\BH^2}\left(\frac{1}{ f_{a}^{2}}\right)M_{211} \quad(\text{attractive self-interaction, for }{\alpha\lesssim0.2}),
\end{equation}
where $ M_{211}$ is the $\, \left|211\right\rangle$ state cloud mass and we consider the attractive self-interaction here (the minus sign). Therefore, we can compare this flat-spacetime estimation~\eqref{freqq} with the result calculated using the method described in Sec.~\ref{met}. The numerically computed result within the Kerr framework is shown in red line in Fig.~\ref{figgw1}. 
It can be seen that, when $ \alpha$ is significantly smaller than unity, there is a good agreement between the numerical results and the analytical estimations.  The red line (numerical result)  in Fig.~\ref{figgw1} approximately follows a trend scaled by $\alpha$ to the fourth power. However, in the regime of $\alpha\gtrsim0.2 $, the analytical result~Eq.~\eqref{freqq} might not be a good estimation (dashed line part) and the non-relativistic approximation used in the analytical method becomes inapplicable, as has been studied in Ref.~\cite{Baryakhtar:2020gao}. 
Similarly,  in the non-relativistic limit\footnote{In the large-$\alpha$ regime, Ref.~\cite{May:2024npn} gives a more careful treatment for the self-gravity effects.},  the self-gravity correction can be estimated by 
\begin{equation}\label{freqg}\varDelta \omega_{211}= -0.19\left.\frac{\alpha^3}{M_\BH^2}\right.M_{211}\quad(\text{self-gravity, for }{\alpha\lesssim0.2}).\end{equation}
It can be seen from Eq.~\eqref{freqq} and Eq.~\eqref{freqg} that, in the small-$\alpha$ regime, the self-interaction correction 
could become dominant over the  self-gravity  effect when the  self-interaction coupling is sufficiently large.  

Moreover, the boson cloud annihilation generates the GWs with frequency twice the boson frequency. 
During the boson cloud evolution,   such as the depletion of cloud mass through annihilation, the  boson frequency can vary with time, which results in a non-zero GW frequency drift in the signals sourced from the cloud.  Those imprints on GWs are perhaps observable in  continuous wave searches. 
The  frequency of GWs from the boson cloud when considering the self-interaction is  
\begin{equation}\begin{aligned}\label{fgw}
{f}_{\rm GW} =& {f}_{\rm GW,\ free} + \frac{\varDelta \omega_{211}}{ \pi}  \,\,,
\end{aligned}\end{equation}
where ${f}_{\rm GW,\ free} $ is GW frequency for the  boson cloud without self-interaction effect. \revt{The additional term in Eq.~\eqref{fgw} contributes to the non-zero GW frequency drift, because the time derivative of cloud mass $\dot{M}_{211}$ is non-zero. 
The depletion rate of cloud is largest when the occupation number reaches the upper bound. That corresponds to the stage when the boson cloud has saturated. After the cloud saturation, the cloud slowly loses its mass by radiating continuous GWs. 
Last but not the  least, from the perspective of GW detection, the frequency shift induced by the self-interaction effects could be degenerate with other parameters, such as boson mass and BH mass. Based on the previous analysis, the amplitude of frequency shift brought by the self-interaction depends on both the self-interaction scale and the occupation number of cloud. Thus, in contrast to the frequency shift caused by other parameters, the frequency shift induced by the self-interaction effects varies differently during the evolution of boson cloud. This characteristic can be conducive to distinguishing the signal of self-interacting boson cloud in the detection.   
}

\section{\label{csq}Observational consequences on superradiance with self-interaction}
In this section, we study further the  self-interaction's observational consequences on the gravitational atom and boson cloud. We focus on two direct consequences of superradiant cloud formation. First, since the BH will spin down during the superradiance process, the spin measurements on the astrophysical BHs are usually used to detect or constrain the bosons with particular mass. Also, the boson cloud will emit GWs through varied processes, which can be used to explore the properties of the boson cloud and the gravitational atom.  
\subsection{Relaxing constraints on scalar boson models}
The direct implication of the self-interaction on the formation of boson cloud is that the cloud growth ceases so long as the occupation number of bound state reaches a critical value $\widetilde{N}_{c}$, which is much smaller than the maximum occupation number ${N}_{c}$ calculated when no self-interaction effect is included. That also means the boson cloud of particular state can extract less angular momenta from the  BH than expected. The BH spin-down would pause  earlier before its spin reached the corresponding critical bound for the superradiance of the free field particle. Moreover, the BH spin-down rate is also smaller because of the self-interaction effect. What we can interpret from these is the fast-spinning astrophysical BH can live long with a high dimensionless spin even though the superradiance once happens, if we assume the bosons have a moderate quartic self-interaction. That means our analysis about the self-interaction effect on the superradiance process can relax the existed constraints on the boson mass range obtained through the astrophysical BH spin measurements.   

\begin{table}[b]  
\centering     
\renewcommand{\arraystretch}{1.4}
\begin{tabular}{>{\centering\arraybackslash}m{0.1\linewidth}|>{\centering\arraybackslash}p{0.25\linewidth}|>{\centering\arraybackslash}p{0.25\linewidth}|>{\centering\arraybackslash}p{0.25\linewidth}}
\multicolumn{4}{c}{\vspace{6pt}{\fontsize{14}{30}\selectfont $\chi_i=0.95,\,M_\BH = 40~{\rm M_\odot},\,m_a=2\times10^{-13}~\mathrm{eV}$}}\\
\hline \hline
$f_{a}$&   $M_\mathrm{Pl}$&   $10^{19}~\mathrm{GeV}$&  $10^{18}~\mathrm{GeV}$\\
\hline
\vspace{0pt}$\widetilde{N}_{c}$&   $ 2.6\times10^{76}$&   $\sim 10^{76}$&  $\sim 10^{74}$\\
\hline 
\vspace{3pt}
 $\dfrac{\widetilde{N}_{c}}{{N}_{c}}$& $ 2.8\ $‰& 1.9~‰&$1.9\times10^{-5}$~\\[1ex]
\hline
$ \varDelta \chi$ &   $2.0\times10^{-3}$&   $1.3\times10^{-3}$&  $1.3\times10^{-5}$\\
\hline \hline
\end{tabular}
\caption{The BH spin difference after the cloud growth ceases for different self-interaction strength $f_{a}$. }  
\label{tabsd2}
\end{table}
\begin{table}[h]  
\centering     
\renewcommand{\arraystretch}{1.4}
\begin{tabular}{>{\centering\arraybackslash}m{0.12\linewidth}|>{\centering\arraybackslash}p{0.25\linewidth}|>{\centering\arraybackslash}p{0.25\linewidth}|>{\centering\arraybackslash}p{0.25\linewidth}}
\multicolumn{4}{c}{\vspace{6pt}{\fontsize{14}{30}\selectfont $\chi_i=0.95,\,m_a=2\times10^{-13}~\mathrm{eV},\,f_{a} = 10^{19}~\mathrm{GeV}$}}\\
\hline \hline
$M_\BH~[{\rm M_\odot}]$&   40&   60&  100\\
\hline
\vspace{0pt}$\widetilde{N}_{c}$&   $\sim 10^{76}$&   $\sim 10^{77}$&  $ 7.8\times10^{77}$\\
\hline 
\vspace{3pt}
 $\dfrac{\widetilde{N}_{c}}{{N}_{c}}$& 1.9~‰& 5.7~‰&2.3~\%\\[1ex]
\hline
$ \varDelta \chi$ &   $1.3\times10^{-3}$&   $3.4\times10^{-3}$&  $9.3\times10^{-3}$\\
\hline \hline
\end{tabular}
\caption{Same as Table~\ref{tabsd2}, but for different BH mass $M_\BH$ (with a fixed self-interaction strength). }  
\label{tabsd4}
\end{table}

\revt{For the boson cloud growth with the self-interaction, the exponential growth rate of the  dominant state is determined by not only the fine-structure constant $\alpha$, the boson mass $m_{a}$ and BH spin $\chi$, but also the self-interaction scale $f_{a}$ and the occupation number of the state $N(t)$. These functional dependences have already been explored in Sec.~\ref{rts}. We can define the normalized growth rate of boson cloud $\omega_\text{I}={\text{Im}(\omega_a)}/{m_{a}}$ , where the growth rate is normalized by the boson mass and becomes dimensionless in natural units. As the occupation number $N(t)$ increases during the cloud growth, the self-interaction effects  become more significant, leading to progressive suppression of  the growth rate $\omega_\text{I}$ (cf. Fig.~\ref{figonac}). Letting the growth rate approaches zero, i.e. 
\begin{equation}
    \omega_\text{I} ( N,\ f_{a} ,\  \alpha ,\ \chi )=0\,\,,
\end{equation}
we can obtain the critical value of occupation number, $\widetilde{N}_{c}$. When $N=\widetilde{N}_{c}$, the cloud growth ceases, and we can estimate the BH spin difference through the conservation of angular momentum: 
\begin{equation}
    \varDelta \chi=\frac{m\widetilde{N}_{c}}{M_\BH^2}.
\end{equation}
Note that $m$ is the magnetic quantum number of bound-state boson. }

\revt{In contrast, if there is not scalar self-interaction, the boson cloud could saturated and the final BH spin in that stage was generally determined by the superradiance condition~\eqref{SR}. Thus, the BH spin difference for free-field case is 
\begin{equation}
    \varDelta \chi_{\rm  free}=\left|\chi_{f}^{}-\chi_{i}\right|,
\end{equation}
where $\chi_{f}$ is BH spin after boson cloud formation and $\chi_{i}$ is initial BH spin. }

We take the boson with mass $m_{a}=2\times10^{-13}~\mathrm{eV}$ and an isolated astrophysical BH $M_\BH = 40~{\rm M_\odot}$ as the example. Assuming the initial dimensionless spin $\chi_{i}=0.95$ after the BH's birth, we concentrate on the $\, \left|211\right\rangle$ state cloud growth, which has the fastest superradiance rate among all states. The BH is isolated from other celestial objects, so the angular momentum can only be transferred from BH to the cloud and keep conservative. When there is not self-interaction in bosons, the $\, \left|211\right\rangle$ cloud will take $\sim180$ e-folding time to grow to maximum number approximately, and the spin difference is $\sim\mathcal{O}(0.1)$ for the BH spin-down. However, with the existence of self-interaction, the BH spin difference after the cloud stops to grow will decrease. We consider the quartic self-interaction with $f_{a}=10^{19},10^{18}~\mathrm{GeV}$ and Planck energy scale, respectively, and calculate the corresponding BH spin difference. The results are shown in Table~\ref{tabsd2}. We find that the difference in BH spin after the $\, \left|211\right\rangle$ state cloud growth ceases is smaller than $2\times10^{-5}$, when the parameter $f_{a}\lesssim10^{18}~\mathrm{GeV}$. In that regime, the occupation number is suppressed by a factor of order $ 10^{4}$ at least. That indicates the BH spin-down is very insignificant, different from the situation for the free field boson. Similarly, we investigate the BH spin difference for BH systems with initial masses of $40$, $60$, and $100~{\rm M_\odot}$, respectively, by fixing the self-interaction strength. These results are shown in Table~\ref{tabsd4}. We can see from Table~\ref{tabsd4} that for a BH with larger initial mass, its spin-down due to the formation of (self-interacting) boson cloud is also larger. Specifically, for the benchmark parameters in Table~\ref{tabsd4}, the spin difference of a $100~{\rm M_\odot}$ BH is $ \varDelta \chi \sim$ 0.01. This value is still 40 times smaller than the result for free-field case. 

Although the analysis above focus on the isolated BH, the result is also applicable for the binary black hole (BBH) system if their inspiral timescale is long enough to undergo the superradiance and the cloud formation. For the $\, \left|211\right\rangle$ state, the cloud growth timescale is usually much smaller than the typical inspiral timescale from the time the BBH is formed to the time the BHs merge. Hence the above result can also describe spin-down processes of individual BHs in the BBH.  

\begin{table}[t]  
\centering     
\renewcommand{\arraystretch}{1.4}
\begin{tabular}{m{0.36\linewidth}<{\centering}|m{0.18\linewidth}<{\centering}|m{0.18\linewidth}<{\centering}}
\hline \hline
&   GW190412&  GW190517\\
\hline
primary BH mass $[{\rm M_\odot}]$ &   $ 27.7$&  $39.2$\\
secondary BH mass $[{\rm M_\odot}]$ & 9.0&24.0\\
 mass of the remnant BH $[{\rm M_\odot}]$& 35.6&60.1\\
primary BH spin& 0.44&0.95\\
spin of the remnant BH &   0.66&   0.87\\
\hline \hline
\end{tabular}
\caption{The parameters of the BBH systems in GW190412 and GW190517.}  
\label{tp1}
\end{table}
\begin{figure}[t]
    \centering
    \includegraphics[width=0.8\linewidth]{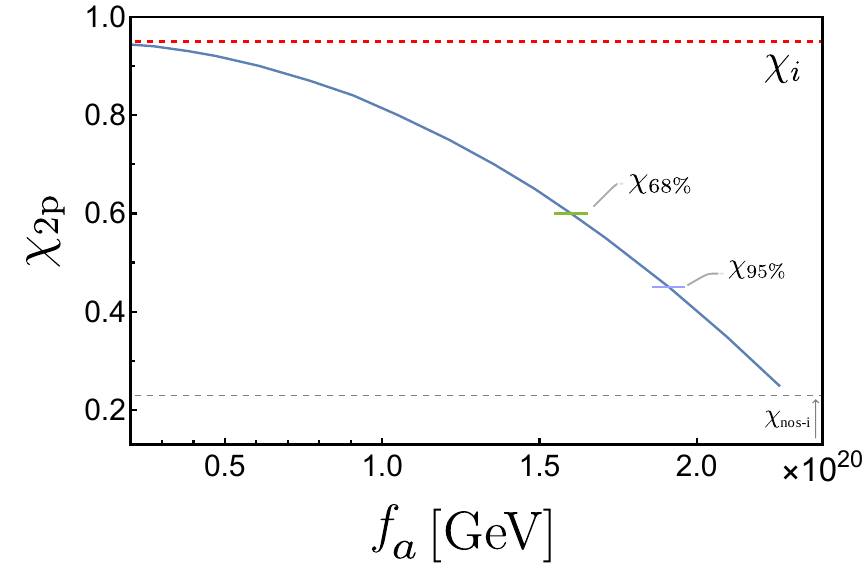}
    \caption{The final BH spin after the cloud growth for the primary BH in GW190517. We find when  $f_{a}\lesssim1.91\times10^{20}~\mathrm{GeV}$, the BH can exist with spin covered by the $95\%$ credible region interpreted from  observation data,  and when  $f_{a}\lesssim1.6\times10^{20}~\mathrm{GeV}$, that  BH can exist having a large spin contained in the $68\%$ credible interval interpreted from GW data. }
    \label{figfsf}
\end{figure}
Currently, LIGO/Virgo/KAGRA  Collaborations have reported more than one hundred GW events, most of which are BBH merger. The BH spin measurement through GW detection are used to constrain the boson mass range~\cite{Ng:2020ruv}. In particular, GW190412~\cite{LIGOScientific:2020stg} and GW190517 give the most stringent constraints on the scalar boson mass among all events in GWTC-2~\cite{Ng:2020ruv}. The parameters of the BHs in these two events are listed in Table~\ref{tp1}. 
If there is long enough time for the bound state boson to grow to a macroscopic cloud before the BHs merge, each BH dimensionless spin should not exceed the critical spin determined by the boson mass and the fastest growing state. That is basically why the BH spin measurement through GW detection can provide the evidence on excluding the boson mass range. However, the individual BHs \textit{can} retain a large spin until they merge even though the boson of the appropriate mass exists, as long as  boson particles have the self-interaction. We take the event GW190517 as the example and assume the initial BH spin when it is formed is equal to the mean value of respective BH spin interpreted from GW data (e.g. for the primary BH in GW190517, it is approximately 0.95). We compute the BH spin difference at the time the $\, \left|211\right\rangle$ cloud growth saturates  as the function of the self-interaction scale $f_{a}$. Fig.~\ref{figfsf} shows the final BH spin after the cloud growth for the primary BH in the binary, and we can figure out that when  $f_{a}\lesssim1.91\times10^{20}~\mathrm{GeV}$, the BH can exist with spin covered by the $95\%$ credible region interpreted from  GW data. Moreover, when  $f_{a}\lesssim1.6\times10^{20}~\mathrm{GeV}$, that  BH can exist having a large spin contained in the $68\%$ credible interval. Similarly for the primary BH in the event GW190412, these scales are $3.3\times10^{20}~\mathrm{GeV}$ and  $2.6\times10^{20}~\mathrm{GeV}$, corresponding to $95\%$ and $68\%$ credible intervals, respectively. In total, we can draw a conclusion that if there exists self-interaction with $f_{a}\leq10^{20}~\mathrm{GeV}$ for the boson mass of $m_{a}=2\times10^{-13}~\mathrm{eV}$, the events GW190412 and GW190517 can not be used to constrain the boson model in that mass range via their BH spin measurements. This self-interaction scale is larger than the scale requirement derived in Ref.~\cite{Ng:2020ruv}, in which the scale $f_{a}$ shall be greater than $10^{14}~\mathrm{GeV}$ so as to safely neglect the bosenova effect in the boson cloud of both events. That implies the self-interaction  effects in the analysis of our work are more common.   
\subsection{Implication for detecting gravitational atom }
The evolution scenario of the BH parameters and the GW signals sourced from the boson cloud through its annihilation could be very different when the bosons have the self-interaction. For a gravitational atom, the  mass and spin of BH are interesting observables in the detection, as well as the strength of GWs emitted by its boson cloud. In the prior section, we have studied the consequences of the self-interaction effects. These consequences can be summarized in the following. (1) The BH will  have a larger spin and mass after the cloud growth terminates, because the self-interaction  brings an extra limitation on the occupation number of bound state and the boson cloud growth will cease when its occupation number reaches $\widetilde{N}_{c}$, which depends on the self-interaction scale $f_{a}$ and BH mass. 
(2) The self-interaction of boson cloud alters the cloud frequency by an amount dependent on the self-interaction strength and the cloud mass. That is because the attractive self-interaction can bind the macroscopic cloud more tightly, and vice versa. The cloud mass decreases slowly when the cloud annihilates and radiates GW. Therefore the GW frequency change during that evolution, and this drift is possibly observable in the detection.  
(3) The lighter boson cloud and smaller occupation number reduce the GW emission power from the cloud. The peak GW strain decreases for the GW signals originated from bound state cloud through bosons annihilation, while the GW emission timescale increases to a certain extent accordingly.  
In order to clearly show the influence of the self-interaction, we numerically compute the evolutions of the BH mass and spin, the occupation number of boson cloud in the gravitational atom,  the cloud's GW strain at the detectors, the GW emission power, and the GW frequency, as the function of time starting from the birth of gravitational atom. We set the benchmark parameters as $M_\text{BH, init}=25~{\rm M_{\odot}},\ \chi_{i}=0.95,\ m_a=10^{-12}~\mathrm{eV \ and}\ f_{a}= 2.5\times10^{18}~\mathrm{GeV}$.   

\begin{figure}[h]
    \centering
    \includegraphics[width=0.27\linewidth]{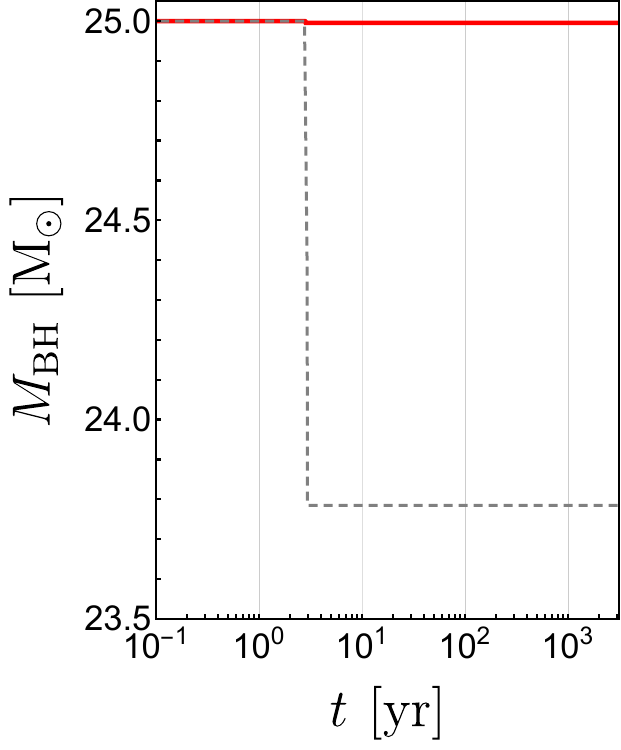}
    \includegraphics[width=0.27\linewidth]{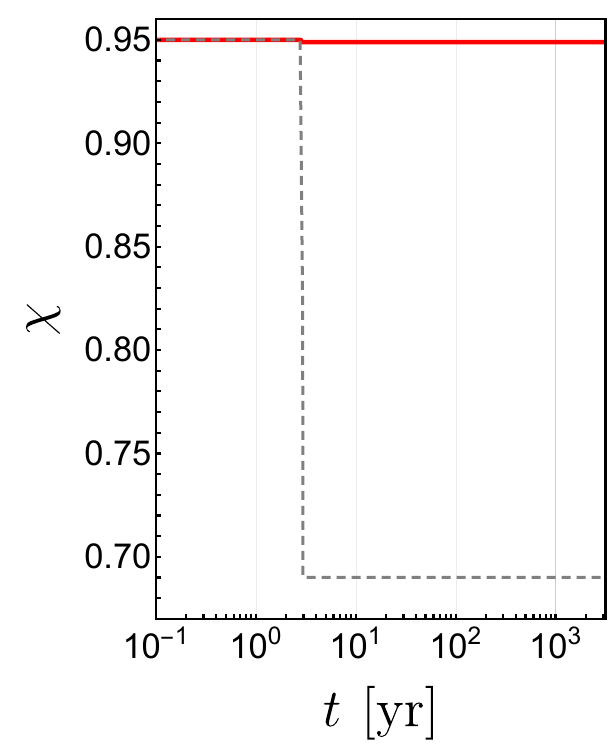}
    \includegraphics[width=0.35\linewidth]{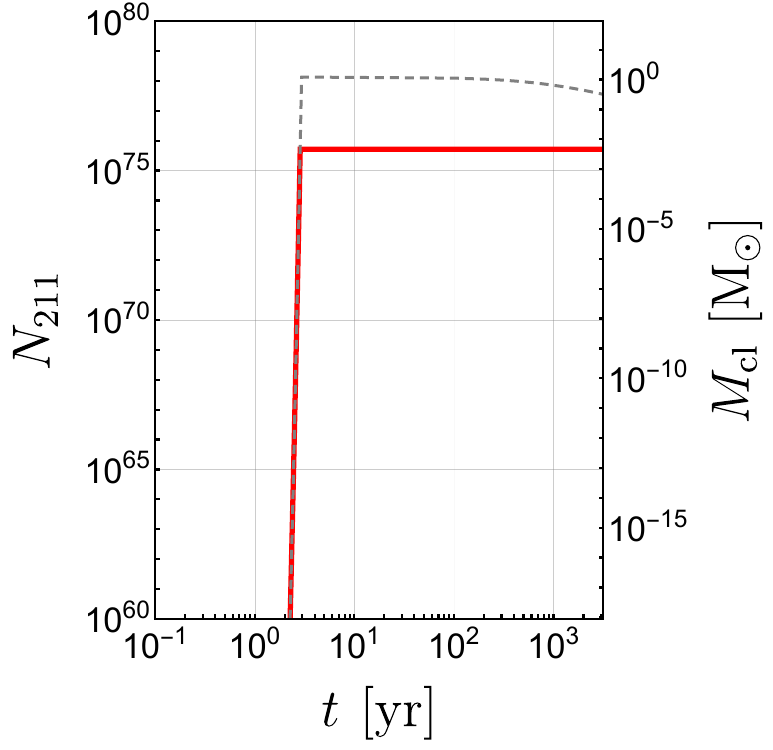}
    \includegraphics[width=0.277\linewidth]{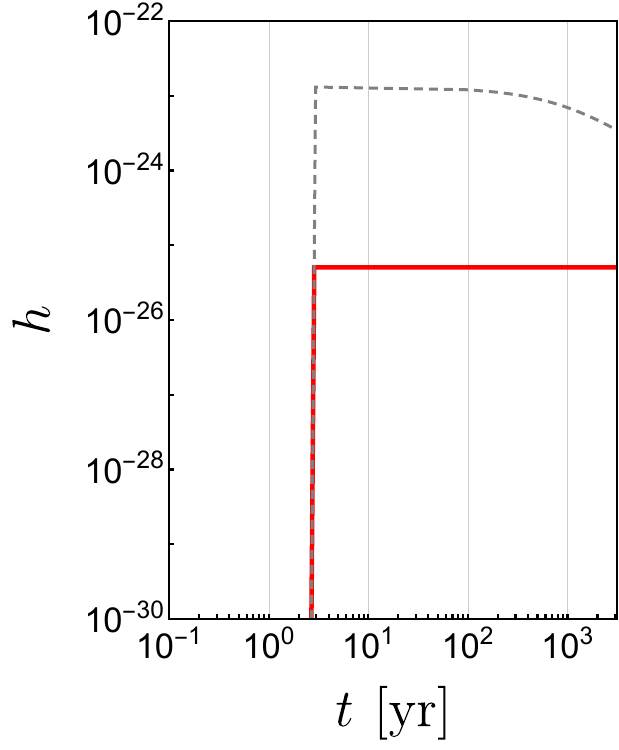}
      \includegraphics[width=0.27\linewidth]{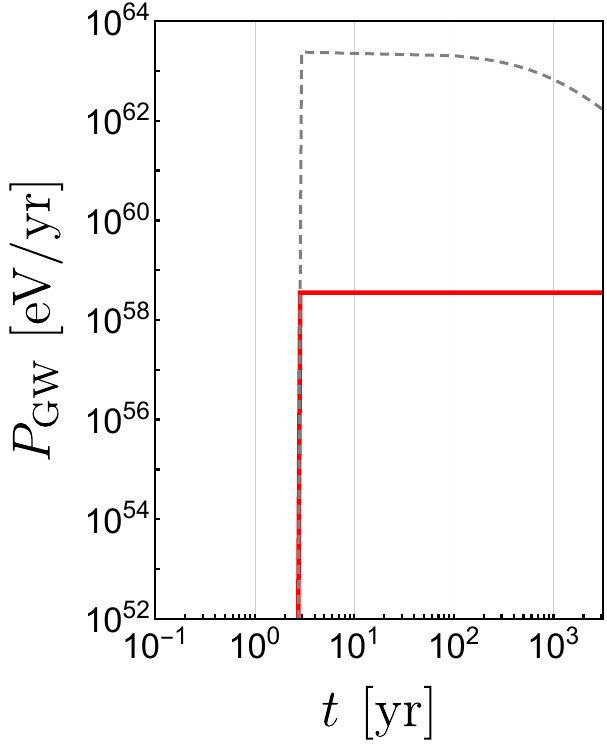}
    \includegraphics[width=0.3\linewidth]{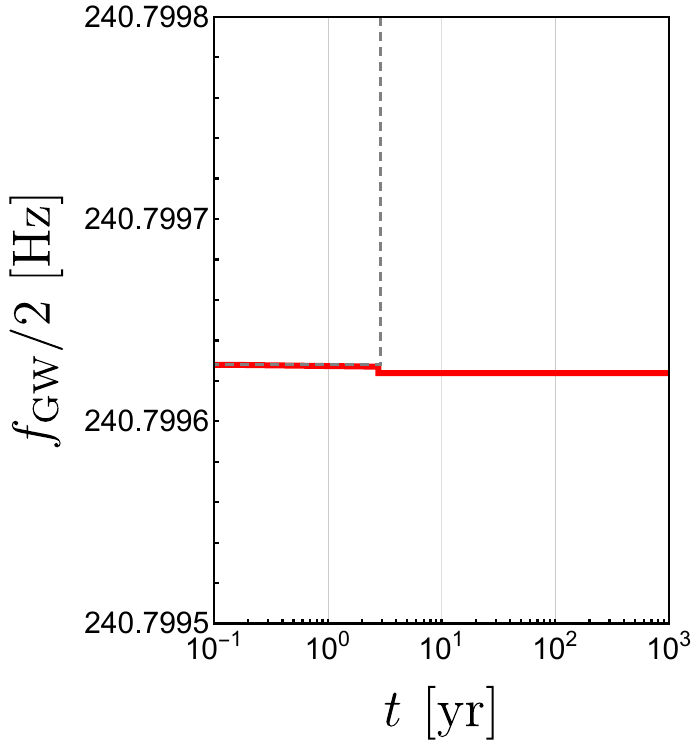}\qquad \qquad
    \caption{Evolution examples of 
    BH mass, 
    BH spin,  ultralight boson cloud occupation number and its mass, cloud's radiated GW strain, GW  emission power, and  GW frequency. The red lines in each subfigure show the evolution of these parameters as the function of time starting from the birth of the gravitational atom. The BH has an initial mass of $M_\text{BH, init}=25~{\rm M_{\odot}}$ and an initial spin of $\chi_{i}=0.95$. The self-interaction scale is  $f_{a}= 2.5\times10^{18}~\mathrm{GeV}$. To make a comparison, the gray dashed lines in each subfigure show the evolution of the parameters if we assume no self-interaction between bosons in the cloud. We assume the luminosity distance of gravitational atom $D_{L}=10~\text{kpc}$ to compute the GW strain at the detectors. The times in the horizontal axes are all rescaled logarithmically. }
    \label{figdet}
\end{figure}
\revt{To clearly show the relation between the GW detection and the gravitational atom system, it is worth to explicate the methods used to derive the GW emission power and GW strain detected. Firstly, in the annihilation process of bosons, the time derivative of  boson cloud occupation number is governed by 
\begin{equation}
    \frac{\text{d}N(t)}{\text{d}t}=-\Gamma_{a} N^2(t) .
\end{equation}
Here, the minus sign represents the depletion of boson cloud occupation number and this differential equation is equivalent to the Eq.~\eqref{ONevo} in Sec.~\ref{mod},  the evolution function of occupation number. 
Then, the emission energy for GWs radiated by boson cloud equals the energy loss of cloud, which means 
\begin{equation}
    P_{\rm GW}=P_{\rm loss}=-\frac{\text{d}N(t)}{\text{d}t}m_{a} .
\end{equation}
Here we approximate the energy of bound-state boson by $m_{a}$. From the perspective of this work, we focus on how the self-interaction effects change the maximum occupation number and lead to the significant reduction of peak GW  strain. We neglect the self-interaction corrections to the annihilation rate and the energy of scalar boson. When calculating the emission power of GWs from boson cloud, those corrections are expected to be subdominant because the amplification of occupation number is largely suppressed. Finally, the (peak) GW strain at the detector can be estimated as 
\begin{equation}
    h\simeq\sqrt{\frac{4 P_{\rm GW}}{5\left(2\pi f_{\rm GW} \right)^2D_{L}^2}} ,
\end{equation}
where $D_{L}$ is the luminosity distance of gravitational atom and the prefactor $4/5$ 
is decided by the configuration of detector. 
}

We demonstrate the evolutions of the gravitational atom's parameters in Fig.~\ref{figdet}.  We assume the luminosity distance of gravitational atom $D_{L}=10~\text{kpc}$ to compute the GW strain at the detectors.  Note that the times in the horizontal axes are all rescaled logarithmically. We only consider the boson cloud of fastest growing state for simplicity and assume the attractive self-interaction. The red lines in Fig.~\ref{figdet} show the parameters' evolution after the gravitational atom is born. As a contrast, we also plot the evolution of those parameters for the free boson in gray dashed lines, in which the self-interaction scale $f_{a}$ is set to be infinite.  We can see from the plots  the cloud growth ceased before the original saturation and this turn-off happened  earlier. The BH mass and spin only decreased slightly and then nearly remained  unchanged. The magnitude of boson cloud was much smaller comparing to the boson of free field. As a result, the GW strain was suppressed to a large extent. Although the boson cloud was radiating GWs, its occupation number did not decline after the time $t>3~\text{years}$. The reason for that is the  BH with a large spin can continuously replenish the cloud through superradiance when the bosons in the cloud annihilated. On the whole, the (peak) GW strain became smaller because of the self-interaction. The smaller GW strain and the frequency drift (not shown in the plots) make it more difficult to discover the signals in the GW searches.   

One important insight those charts in Fig.~\ref{figdet} reveal is that, at the time the boson cloud is saturated, the frequency of GWs emitted by the cloud changes dramatically. This change stems from the energy shift of the bound-state bosons owing to both the self-interaction effects of boson cloud and the rapid mass depletion of BH. To be specific, the self-interaction effects would alter the BH mass depletion and the frequency eigenvalue of boson cloud during the evolution of gravitational atom. Thus, the observation and GW frequency measurement of continuous GWs radiated by the newly-formed boson cloud 
could constrain the boson models with self-interactions.  
The detailed studies are left for future work. 

\section{\label{cc}Conclusions and discussions}

The successful detection of GWs has opened new avenues and provided powerful probes for studying ultralight bosonic fields and dark matter. Traditional dark matter searches often rely on its electromagnetic or other interactions with ordinary matter. However, if dark matter consists of ultralight bosonic fields, it can spontaneously form and grow into macroscopic, dense boson clouds around rotating BHs without requiring any interactions beyond gravity. The gravitational atom formed by a rotating BH and its surrounding boson cloud exhibits various observable GW signatures, including modifications to GW emissions from compact binaries~\cite{Xie:2022uvp,Arvanitaki:2016qwi,Hannuksela:2018izj}  and continuous GW radiation from the boson cloud itself~\cite{Arvanitaki:2010sy,Arvanitaki:2014wva,Brito:2017wnc,Brito:2017zvb,East:2018glu,Siemonsen:2022yyf,Yang:2023vwm,Yang:2023aak}.

\revt{Apart from these, whether the ultralight bosons  possess self-interactions   can significantly affect: (i) the formation, saturation, and evolution of the boson cloud; (ii) the evolution of the BH; and (iii) the GW signals from the gravitational atom.} When the self-interaction is sufficiently strong, both the final state of the BH and the mass of the boson cloud can differ substantially from the non self-interacting case.

In this work, we have studied the formation, growth rate, saturation, and bound state energy levels of self-interacting boson clouds during the BH superradiance process within the full Kerr background, while also accounting for the corresponding evolution of the BH's mass and spin. Previous studies on dark matter detection or constraints via superradiance have largely overlooked the self-interaction effects of boson clouds. Here, we numerically solve for the superradiant growth rate of boson clouds with self-interactions around Kerr BHs and quantitatively evaluate how the strength of scalar boson self-interactions influences the growth rate.

Our calculations reveal that, when self-interactions are included, there exists an additional critical boson cloud mass beyond the upper bound imposed by the superradiance condition. Once the boson cloud reaches this critical mass during its growth, the growth rate approaches zero, indicating that the superradiance process will terminate earlier under the influence of self-interactions. The maximum mass of the boson cloud thus depends on the strength of the  self-interaction of ultralight bosons. Furthermore, we numerically compute the real part of the eigenfrequencies of the boson cloud in the gravitational atom and find deviations from the non-interacting case, with the shift depending on both the cloud mass and the self-interaction strength.

Based on these results, we discuss the impact of self-interactions on GWs emitted from boson cloud annihilation in gravitational atoms, including both frequency and amplitude modifications. The predicted frequency drift of the emitted GWs could be detectable by current or future observatories. Moreover, our calculations suggest that, for gravitational atoms composed of scalar boson clouds, the BH may retain a significant fraction of its initial mass and spin even after the dominant superradiant mode terminates if self-interactions are present. This finding could substantially relax current constraints on scalar boson models derived from BH spin measurements. As specific examples, we use GW190412 and GW190517 to demonstrate how self-interaction can alleviate these constraints, providing quantitative estimates of the self-interaction energy scale required to do so. Additionally, our results can help explain several existing experimental measurements of BH spins.

Overall, the conclusions drawn from this work provide valuable insights for future dark matter searches using GW observations and for astrophysical constraints on dark matter properties based on BH observations. In investigating the superradiance of self-interacting scalar bosons around rotating BHs, we solve the Klein-Gordon equation on the Kerr metric and, for the first time in a relativistic framework, apply the continued fraction method to compute the eigenfrequencies of boson clouds with self-interactions in gravitational atoms. This approach allows us to determine the superradiant growth rates of self-interacting bosons for specific energy levels around Kerr BHs and, in principle, to numerically solve for the growth rates at any stage of the boson cloud formation across the entire parameter space of $\alpha$ and $f_a$.
In the limit where the self-interaction scale $f_a$
approaches infinity, our results recover those of the non-interacting case.

Nevertheless, our calculations assume that the boson cloud is dominated by a single energy level, neglecting mode mixing between different states and the influence of higher bound states on the dominant mode. Future work could expand upon this study in several directions. For example, one could consider higher-order self-interaction terms beyond the quartic term, incorporating their effects within the current methodological framework. Additionally, the impact of self-interactions on the annihilation of boson pairs into gravitons within the cloud could be explored in greater detail, providing a more comprehensive understanding of GW signals from gravitational atoms. From a methodological perspective, applying numerical relativity to dynamically model the formation and evolution of self-interacting boson clouds around a BH could yield more precise results for system parameters such as growth rates and validate the methods employed in this study.

\begin{acknowledgments}
We thank Siyu Jiang, Dayun Qiu, Jing Yang, Xiang'en Yang, and Junquan Su for their useful suggestions on the manuscript.
This work was supported by the National Natural Science Foundation of China (NNSFC) under Grants No.~12475111 and No.~12205387.
    
\end{acknowledgments}
\bibliography{reference}

\end{document}